\documentclass{ws-ijmpd}
\begin{document}

\markboth{R. Sharma and B. S. Ratanpal}
{Relativistic stellar model admitting a quadratic equation of state}

%
\catchline{}{}{}{}{}
%

\title{RELATIVISTIC STELLAR MODEL ADMITTING A QUADRATIC EQUATION OF STATE}

\author{R. SHARMA}

\address{Department of Physics, P. D. Women's College, Jalpaiguri 735 101, India.\\
E-mail: rsharma@iucaa.ernet.in}

\author{B. S. RATANPAL}

\address{Department of Applied Mathematics, Faculty of Technology and Engineering,\\ 
The M. S. University of Baroda, Vadodara  390 001, Gujarat, India.\\
E-mail: bharatratanpal@gmail.com}

\maketitle

\begin{history}
\received{Day Month Year}
\revised{Day Month Year}
\end{history}

\begin{abstract}
A class of solutions describing the interior of a static spherically symmetric compact anisotropic star is reported. The analytic solution has been obtained by utilizing the Finch and Skea ({\it Class. Quant. Grav.} {\bf 6} (1989) 467) ansatz for the metric potential $g_{rr}$ which has a clear geometric interpretation for the associated background space-time. Based on physical grounds appropriate bounds on the model parameters have been obtained and it has been shown that the model admits an equation of state (EOS) which is quadratic in nature. 
\end{abstract}


\ccode{PACS numbers:04.20.-q; 04.20.Jb; 04.40.Dg; 12.39.Ba}

\section{Introduction}	
To construct models of relativistic compact stars, it is imperative to know the exact composition and nature of particle interactions at extremely high density regime. If the equation of state (EOS) of the material composition of a compact star is known, one can easily integrate the Tolman-Oppenheimer-Volkoff (TOV) equations to analyze the physical features of the star. The problem is that we still lack reliable information about physics of particle interactions at extremely high density that may be found in the `natural laboratories' of compact astrophysical objects. 

The objective of the present paper is to construct models of equilibrium configurations of relativistic compact objects when no reliable information about the composition and nature of particle interactions are available. This can be achieved by generating exact solutions of Einstein's field equations describing the interior of a static spherically symmetric relativistic star. However, finding exact solutions of Einstein's field equations is extremely difficult due to highly non-linear nature of the governing field equations. Consequently, many simplifying assumptions are often made to tackle the problem.  Since General Relativity provides a mutual correspondence between the material composition of a relativistic star and its associated space-time, we will adopt a geometric approach to deal with such a situation. In this approach, a suitable ansatz for one of the metric potentials with a clear geometric characterization  of the associated space-time metric will be prescribed to determine the other. Such a method was initially proposed by Vaidya and Tikekar\cite{Vaidya}; subsequently the method was utilized by many to generate and analyze physically viable models of compact astrophysical objects (see for example, \refcite{Knutsen,Tikekar,Maharaj,SNB,Sharma01,Sharma02,Sharma03,Sharma04,Sharma05,sk} and references therein). In the present work, we shall utilize the Finch and Skea\cite{Finch} ansatz for the metric potential $g_{rr}$ to determine the unknown metric potential $g_{tt}$ describing the interior space-time of a static spherically symmetric stellar configuration. Note that the $t=constant$ hyper-surface of the background space-time corresponding to the Finch and Skea\cite{Finch} ansatz is paraboloidal in nature\cite{Tikekar2007}. 

In our work, we shall incorporate a general anisotropic term in the stress-energy tensor representing the material composition of the star. We would like to point out here that anisotropic matter is a very exotic choice for compact objects like neutron stars. Nevertheless, in the past, impacts of anisotropic stresses on equilibrium configurations of relativistic stars have been extensively investigated by Bowers and Liang\cite{Bowers} and Herrera and Santos\cite{Herrera1}. Local anisotropy at the interior of an extremely dense object may occur due various factors such as the existence of type 3A super-fluid\cite{Bowers,Ruder,KW}, phase transition\cite{Soko}, presence of electromagnetic field\cite{Ivanov2010}, etc. In \refcite{Iva2010}, it has been shown that influences of shear, electromagnetic field etc. on self-bound systems can be absorbed if the system is considered to be anisotropic, in general. Mathematically, anisotropy provides an extra degree of freedom in our system of equations. Therefore, on top of Finch and Skea ansatz, we shall utilize this freedom to assume a particular pressure profile to solve the system. In the past, a large class of exact solutions corresponding to spherically symmetric anisotropic matter distributions have been found and analyzed (see for example, Ref.~\refcite{Sharma,Bayin,Krori,Maharaj1989,Gokhroo,Mak,Mak2,Mak3}).  Maharaj and Chaisi\cite{Maharaj2006} have prescribed an algorithm to generate anisotropic models from known isotropic solutions. Dev and Gleiser\cite{Dev1,Dev2,Dev3} have studied the effects of anisotropy on the properties of spherically symmetric gravitationally bound objects and also investigated stability of such configurations. It has been shown that if the tangential pressure $p_{\perp}$ is greater than the radial pressure $p_r$ of a stellar configuration, the system becomes more stable. Impact of anisotropy has also been investigated by Ivanov\cite{Ivanov2002}. In an anisotropic stellar model for strange stars developed by Paul {\it et al}\cite{Paul}, it has been shown that the value of the bag constant depends on the anisotropic parameter. For a charged anisotropic stellar model governed by the MIT bag model EOS, Rahaman {\it et al}\cite{Rahaman} have shown that the bag constant depends on the compactness of the star. Making use of the Finch and Skea\cite{Finch} ansatz, Tikekar and Jotania\cite{Tikekar2007} have developed a two parameter family of solutions of Einstein's  field equations and showed the relevance of the class of solutions for the description of strange stars. A core-envelope type model describing a gravitationally bound object with an anisotropic fluid distribution has been obtained in Ref.~\refcite{TRV,TR,TT}.

In our work, following Finch and Skea\cite{Finch} prescription, we have constructed a non-singular anisotropic stellar model satisfying all the necessary conditions of a realistic compact star. Based on physical grounds, we have prescribed bounds on the model parameters and generated the relevant EOS for the system. An interesting feature of our model is that the solution admits a quadratic EOS. It is often very difficult to generate an EOS ($p = p(\rho)$) from known solutions of Einstein's field equations due to mathematically involved expressions. In fact, in most of the models involving an EOS, the EOS is prescribed a priori to generate the solutions. For example, Sharma and Maharaj\cite{Sharma} have obtained an analytic solution for compact anisotropic stars where a linear EOS was assumed. Thirukkanesh and Maharaj\cite{Thiru} have assumed a linear EOS to obtain solutions of an anisotropic fluid distribution. Feroze and Siddiqui\cite{Feroze} and Maharaj and Takisa\cite{MT} have separately utilized a quadratic EOS to generate solutions for static anisotropic spherically symmetric charged distributions. A general approach to deal with anisotropic charged fluid systems admitting a linear or non-linear EOS have been discussed by Varela {\em et al}\cite{Varela}. In our model, we do not prescribe the EOS; rather the solution imposes a constraint on the EOS corresponding to the material composition of the highly dense system.

The paper has been organized as follows. In Section $2$, the relevant field equations describing a gravitationally bound spherically symmetric anisotropic stellar configuration in equilibrium have been laid down. We have solved the system of equations in Section $3$ and analyzed bounds on the model parameters based on physical grounds. Physical features of the model have been discussed in Section $4$. We have also generated an approximated EOS in this section which has been found to be quadratic in nature. In Section $5$, we have concluded by pointing out some interesting features of our model.

\section{Field equations}
We write the interior space-time of a static spherically symmetric stellar configuration in the standard form
\begin{equation}
ds^{2} = e^{\nu(r)}dt^{2}-e^{\lambda(r)}dr^2-r^2\left(d\theta^2+\sin^2\theta d\phi^2\right),\label{eq1}
\end{equation}
where $\nu(r)$ and $\lambda(r)$ are yet to be determined. We assume that the material composition of the configuration is anisotropic in nature and accordingly we write the energy-momentum tensor in the form
\begin{equation}
T_{ij} = \left(\rho + p\right)u_{i}u_{j} - p g_{ij} + \pi_{ij},\label{eq2}
\end{equation}
where $\rho$ and $p$ represent energy-density and isotropic pressure of the system and $u^i$ is the $4$-velocity of fluid. The anisotropic stress-tensor $\pi_{ij}$ is assumed to be of the form
\begin{equation}
\pi_{ij} = \sqrt{3}S\left[C_{i}C_{j}-\frac{1}{3}\left(u_{i}u_{j}-g_{ij}\right)\right],\label{eq3}
\end{equation}
where $S = S(r)$ denotes the magnitude of anisotropy and $C^{i} = \left(0,-e^{-\lambda/2},0,0\right)$ is a radially directed vector. We calculate the non-vanishing components of the energy-momentum tensor as 
\begin{equation}
T^{0}_{0} = \rho,~~~T^{1}_{1} = -\left(p+\frac{2S}{\sqrt{3}}\right),~~~~T^{2}_{2} = T^{3}_{3} = -\left(p-\frac{S}{\sqrt{3}}\right),\label{eq4}
\end{equation}
which implies that the radial pressure and the tangential pressure will take the form
\begin{eqnarray}
p_{r} &=& p + \frac{2S}{\sqrt{3}},\label{eq5}\\
p_{\perp} &=& p - \frac{S}{\sqrt{3}},\label{eq6}
\end{eqnarray}
respectively. Therefore, magnitude of the anisotropy is obtained as
\begin{equation}
p_{r} - p_{\perp} = \sqrt{3} S.\label{eq7}
\end{equation}
The Einstein's field equations corresponding to the space-time metric (\ref{eq1}) and the energy-momentum tensor (\ref{eq2}) are obtained as (in relativistic units with $G = c = 1$)
\begin{eqnarray}
8\pi\rho &=& \frac{1}{r^2}-e^{-\lambda}\left(\frac{1}{r^2}-\frac{\lambda'}{r}\right),\label{eq8}\\
8\pi p_{r} &=& e^{-\lambda}\left(\frac{1}{r^2}+\frac{\nu'}{r}\right)-\frac{1}{r^2},\label{eq9}\\
8\pi p_{\perp} &=& \frac{e^{-\lambda}}{4}\left[2\nu''+\left(\nu'-\lambda'\right)\left(\nu'+\frac{2}{r}\right)\right].\label{eq10}
\end{eqnarray}
Defining the mass within a radius $r$ as
\begin{equation}
m(r) = \frac{1}{2}\int_{0}^{r} {\tilde{r}}^{2}\rho(\tilde{r})d\tilde{r}.\label{eq11}
\end{equation}
we rewrite the field equations (\ref{eq8})-(\ref{eq10}) in the form
\begin{eqnarray}
e^{-\lambda} &=& 1-\frac{2m}{r},\label{eq12}\\
r\left(r-2m\right)\nu' &=& 8\pi p_{r}r^3+2m,\label{eq13}\\
\left(8\pi\rho+8\pi p_{r}\right)\nu'+2(8\pi p_{r}') &=& -\frac{4}{r}\left(8\pi\sqrt{3}S\right).\label{eq14}
\end{eqnarray}

\section{Interior solution}
To solve the system of equations (\ref{eq12}) - (\ref{eq14}), we make use of the Finch and Skea\cite{Finch} ansatz for the metric potential $g_{rr}$ as
\begin{equation}
e^{\lambda(r)} = 1+\frac{r^2}{R^2},\label{eq15}
\end{equation}
where $R$ is a curvature parameter. The ansatz (\ref{eq15}) has a geometric interpretation and was previously found to generate solutions for compact stellar objects\cite{Tikekar2007}. Note that the $t = constant$ hyper-surface of the metric (\ref{eq1}) for the ansatz (\ref{eq15}) represents a paraboloidal space-time immersed in $4$-Euclidean space-time. 

The energy density and mass function are then obtained as
\begin{eqnarray}
8\pi\rho &=& \frac{3+\frac{r^2}{R^2}}{R^2\left(1+\frac{r^2}{R^2}\right)^2},\label{eq16}\\
m(r) &=& \frac{r^3}{2R^2\left(1+\frac{r^2}{R^2}\right)}.\label{eq17}
\end{eqnarray}
Combining Eqs.~(\ref{eq13}) and (\ref{eq17}), we get
\begin{equation}
\nu' = \left(8\pi p_{r}\right)r\left(1+\frac{r^2}{R^2}\right)+\frac{r}{R^2}.\label{eq18}
\end{equation}
To integrate Eq.~(\ref{eq18}), we choose $8\pi p_{r}$ in the form
\begin{equation}
8\pi p_{r} = \frac{p_{0}\left(1-\frac{r^2}{R^2}\right)}{R^2\left(1+\frac{r^2}{R^2}\right)^2},\label{eq19}
\end{equation}
where $p_0 > 0$ is a parameter such that $\frac{p_{0}}{R^2}$ denotes the central pressure. The particular form of the radial pressure profile assumed here is reasonable due to the following facts:
\begin{enumerate}
\item Differentiation of Eq.~(\ref{eq19}) yields
\begin{equation}
\frac{dp_r}{dr} = - p_0 \frac{r}{2\pi (r^2+R^2)^2}.\label{eq20}
\end{equation}
For $p_0 > 0$, Eq.~(\ref{eq20}) implies that $dp_r/dr < 0$, i.e., the radial pressure is a decreasing function of the radial parameter $r$.
At a finite radial distance $r = R$ the radial pressure vanishes which is an essential criterion for the construction of a realistic compact star. The curvature parameter $R$ is then identified as the radius of the star. 
\item The particular choice (\ref{eq19}) makes Eq.~(\ref{eq18}) integrable. 
\end{enumerate}
Substituting Eq.~(\ref{eq19}) in Eq.~(\ref{eq18}), we obtain
\begin{equation}
\nu' = \frac{2p_{0}r}{R^2\left(1+\frac{r^2}{R^2}\right)}+\left(1-p_{0}\right)\frac{r}{R^{2}},\label{eq21}
\end{equation}
which is integrable and yields
\begin{equation}
e^{\nu} = C\left(1+\frac{r^2}{R^2}\right)^{p_{0}}e^{(1-p_{0})r^2/2R^{2}},\label{eq22}
\end{equation}
where $C$ is a constant of integration. Thus, the interior space-time of the configuration takes the form
\begin{eqnarray}
ds^2 &=& C\left(1+\frac{r^2}{R^2}\right)^{p_{0}}e^{(1-p_{0})r^2/2R^{2}}dt^2-\left(1+\frac{r^2}{R^2}\right)dr^2 \nonumber \\
&& -r^2\left(d\theta^2+\sin^2\theta d\phi^2\right),\label{eq23}
\end{eqnarray}
which is non-singular at $r=0$.  

Making use of Eqs.~(\ref{eq14}), (\ref{eq16}), (\ref{eq19}) and (\ref{eq21}), we determine the anisotropy as
\begin{eqnarray}
8\pi\sqrt{3}S = -\frac{\frac{r^2}{R^2}}{4R^{2}\left(1+\frac{r^2}{R^2}\right)^3}\left[\left((3+p_{0})+(1-p_{0})\frac{r^2}{R^2}\right)\right. \nonumber \\
\times\left.\left(2p_{0}+(1-p_{0})\left(1+\frac{r^2}{R^2}\right)\right)+4p_{0}\left(\frac{r^2}{R^2}-3\right)\right]. \label{eq24}
\end{eqnarray}
Note that anisotropy vanishes at the centre ($r=0$) as expected. The tangential pressure takes the form 
\begin{equation}
8\pi p_{\perp} = 8\pi p_{r} - 8\pi \sqrt{3}S = \frac{4p_{0}\left(1-\frac{r^{4}}{R^{4}}\right)+\frac{r^{2}}{R^{2}}f(r, p_0,R)}{4R^{2}\left(1+\frac{r^{2}}{R^{2}}\right)^3},\label{eq25}
\end{equation}
where,
\begin{equation}
f(r,p_0,R) =  \left[\left(3+p_{0}+\left(1-p_{0}\right)\frac{r^{2}}{R^{2}}\right)\left(2p_{0}+\left(1-p_{0}\right)\left(1+\frac{r^{2}}{R^{2}}\right)\right)+4p_{0}\left(\frac{r^{2}}{R^{2}}-3\right)\right].\nonumber
\end{equation}

\subsection{Determination of the model parameters}
Our model has three independent parameters, namely, $p_0$, $C$ and $R$. The requirement that the interior metric (\ref{eq23}) should be matched to the Schwarzschild exterior metric
\begin{equation}
ds^2 = \left(1-\frac{2M}{r}\right)dt^2-\left(1-\frac{2M}{r}\right)^{-1}dr^2 - r^2\left(d\theta^2+\sin^2\theta d\phi^2\right),\label{eq26}
\end{equation}
across the boundary $r=R$ of the star together with the condition that the radial pressure should vanish at the surface ($p_{r}(r=R) = 0$) help us to determine these constants. Note that the form of the radial pressure profile is such that the condition  $p_{r}(r=R) = 0$ itself becomes the definition of the radius $R$ of the star in this construction. Matching the relevant metric coefficients across the boundary $R$ then yields 
\begin{eqnarray}
R  &=& 4 M,\label{eq27}\\
C &=& \frac{e^{-(1-p_{0})/2}}{2^{p_{0} + 1}},\label{eq28}
\end{eqnarray}
where $M$ is the total mass enclosed within the boundary surface $R$. If radius $R$ is known, Eq.~(\ref{eq27}) can be utilized to determine the total mass $M$ of the star and vice-versa. For a given value of $p_0$, Eq.~(\ref{eq28}) determines $C$. Note that in this model, $p_0/R^2$ corresponds to the central pressure and, therefore, for a given mass ($M$) or radius ($R$), if the central pressure is specified the system is completely determined.  

\subsection{Bounds on the model parameters}
Following Finch and Skea\cite{Finch2} and Delgaty and Lake\cite{Delgaty}, we impose the following conditions on our system so that it becomes a physically acceptable model.
\begin{romanlist}[(i)]
\item $\rho(r),~p_{r}(r),~p_{\perp}(r) \geq 0 $, for $ 0 \leq r \leq R$.
\item $\rho-p_{r}-2p_{\perp} \geq 0$, for $ 0 \leq r \leq R$.
\item $\frac{d\rho}{dr},~\frac{dp_{r}}{dr},~\frac{dp_{\perp}}{dr} < 0$, for $0 \leq r \leq R$.
\item $0 \leq \frac{dp_{r}}{d\rho} \leq 1$; $0 \leq \frac{dp_{\perp}}{d\rho} \leq 1$, for $0 \leq r \leq R$.
\end{romanlist}
Note that the requirements (i) and (ii) imply that the weak and dominant energy conditions are satisfied. Condition (iii) ensures regular behaviour of the energy density and two pressures while condition (iv) is invoked to ensure that the sound speed be causal. In addition, for regularity, we demand that the anisotropy should vanish at the centre, i.e., $p_{r} = p_{\perp}$ at $r = 0$. From Eq.~(\ref{eq24}), we note that the anisotropy vanishes at $r=0$ and $S(r) > 0$ for $0 < r < R$. Interestingly, for a particular choice $p_0 = 1$, the anisotropy also vanishes at the boundary $r=R$ in this construction. From Eq.~(\ref{eq16}), it is obvious that $\rho > 0$, and
\begin{equation}\label{eq29}
8\pi\frac{d\rho}{dr} = \frac{-2r\left(5+\frac{r^{2}}{R^{2}}\right)}{R^{4}\left(1+\frac{r^{2}}{R^{2}}\right)^{3}},
\end{equation}
decreases radially outward. We have already stated that $p_0/R^2$ corresponds to the central pressure which implies that $p_0 > 0$. From Eq.~(\ref{eq25}), it can be shown that for $p_{\perp} > 0$, we must have $p_0 < 1$. Thus, a bound on $p_{0}$ is obtained as
\begin{equation}\label{eq30}
0 < p_{0} \leq 1.
\end{equation} 
To obtain a more stringent bound on $p_0$, we evaluate
\begin{equation}\label{eq31}	
8\pi\frac{dp_{\perp}}{dr} = \frac{r\left[\left(3-20p_{0}+p_{0}^2\right)+\left(2+12p_{0}-6p_{0}^2\right)\frac{r^{2}}{R^{2}}+\left(-1-4p_{0}+5p_{0}^{2}\right)\frac{r^{4}}{R^{4}}\right]}{2R^{4}\left(1+\frac{r^{2}}{R^{2}}\right)^4},
\end{equation}
at two different points. At the centre of the star $(r=0)$
\begin{equation}\label{eq32}
\left(8\pi\frac{dp_{\perp}}{dr}\right)_{(r=0)} = 0,
\end{equation}
and the boundary of the star ($r=R$), it takes the form
\begin{equation}\label{eq33}
\left(8\pi\frac{dp_{\perp}}{dr}\right)_{(r=R)} = \frac{1-3p_{0}}{8R^{3}},
\end{equation}
which will be negative if $p_{0} > \frac{1}{3}$. Therefore, a more stringent bound on the parameter $p_{0}$ is obtained as
\begin{equation}\label{eq34}
\frac{1}{3}<p_{0}\leq 1.
\end{equation}

To verify whether the bound on $p_0$ satisfies the causality condition $0 < \frac{dp_r}{d\rho} < 1$, we combine Eqs.~(\ref{eq20}) and (\ref{eq29}), to yield
\begin{equation}\label{eq35}
\frac{dp_{r}}{d\rho} = \frac{p_{0}\left(3-\frac{r^{2}}{R^{2}}\right)}{5+\frac{r^{2}}{R^{2}}}.
\end{equation}
Now, at the centre of the star $(r=0)$, $\frac{dp_{r}}{d\rho} < 1$ if the condition $p_{0} < 1.6667$ is satisfied and at the boundary of the star $(r=R)$, $\frac{dp_{r}}{d\rho} < 1$ if the condition $p_{0} < 3$ is satisfied. Both these restrictions are consistent with the requirement given in (\ref{eq34}).

Similarly, we evaluate
\begin{equation}\label{eq36}
\frac{dp_{\perp}}{d\rho} = \frac{\left(-3+20p_{0}-p_{0}^2 \right)+\left(-2-12p_{0}+6p_{0}^2 \right)\frac{r^{2}}{R^{2}}+\left(1+4p_{0}-5p_{0}^{2} \right)\frac{r^{4}}{R^{4}}}{4\left(1+\frac{r^{2}}{R^{2}} \right)\left(5+\frac{r^{2}}{R^{2}} \right)},
\end{equation}
throughout the star. At the centre $\left(r=0 \right)$, the requirement $\frac{dp_{\perp}}{d\rho} < 1$ puts a constraint on $p_0$ such that $p_{0} < 1.2250$. At the boundary of the star the corresponding requirement is given by  $p_{0} < 4.3333$. Both these requirements are also consistent with the bound $\frac{1}{3} < p_{0}\leq 1$.

\subsection{Stability}
We now investigate the bound on the model parameters based on stability. To check stability of our model, we shall use Herrera's\cite{Herrera2} overtuning technique which states that the region for which radial speed of sound is greater than the transverse speed of sound is a potentially stable region. The radial and tangential sound speeds in our model are obtained as
\begin{eqnarray}
v_{sr}^{2} &=& \frac{dp_{r}}{d\rho} = \frac{p_{0}\left(3-\frac{r^{2}}{R^{2}} \right)}{5+\frac{r^{2}}{R^{2}}},\label{eq37}\\
v_{st}^{2} &=& \frac{dp_{\perp}}{d\rho} = \frac{\left(-3+20p_{0}-p_{0}^2 \right)+\left(-2-12p_{0}+6p_{0}^2 \right)\frac{r^{2}}{R^{2}}+\left(1+4p_{0}-5p_{0}^{2} \right)\frac{r^{4}}{R^{4}}}{4\left(1+\frac{r^{2}}{R^{2}} \right)\left(5+\frac{r^{2}}{R^{2}}\right)}.\label{eq38}
\end{eqnarray}
Herrera's\cite{Herrera2} prescription demands that we must have $v_{st}^{2}-v_{sr}^{2} < 0$ throughout the star. Now, at the centre of the star
\begin{equation}\label{eq39}
\left(v_{st}^{2}-v_{sr}^{2} \right)_{\left(r=0 \right)} = \frac{-3+8p_{0}-p_{0}^{2}}{20}.
\end{equation}
For $\left(v_{st}^{2}-v_{sr}^{2} \right)_{\left(r=0 \right)} < 0$, it is required that $-3+8p_{0}-p_{0}^2 < 0$, i.e., $p_{0} < 0.3944$. 
At the boundary of the star, we have 
\begin{equation}\label{eq40}
\left(v_{st}^{2} - v_{sr}^{2} \right)_{\left(r=R \right)} = -\frac{\left(1+p_{0} \right)}{12},
\end{equation}
which is obviously negative for $\frac{1}{3} <p_{0} <0.3944$. Therefore, our model is physically reasonable and stable if the following bound is imposed: $\frac{1}{3} < p_{0} < 0.3944$.

\section{Physical analysis}
\begin{figure}[pb]
\label{fig:1}
\centerline{\psfig{file=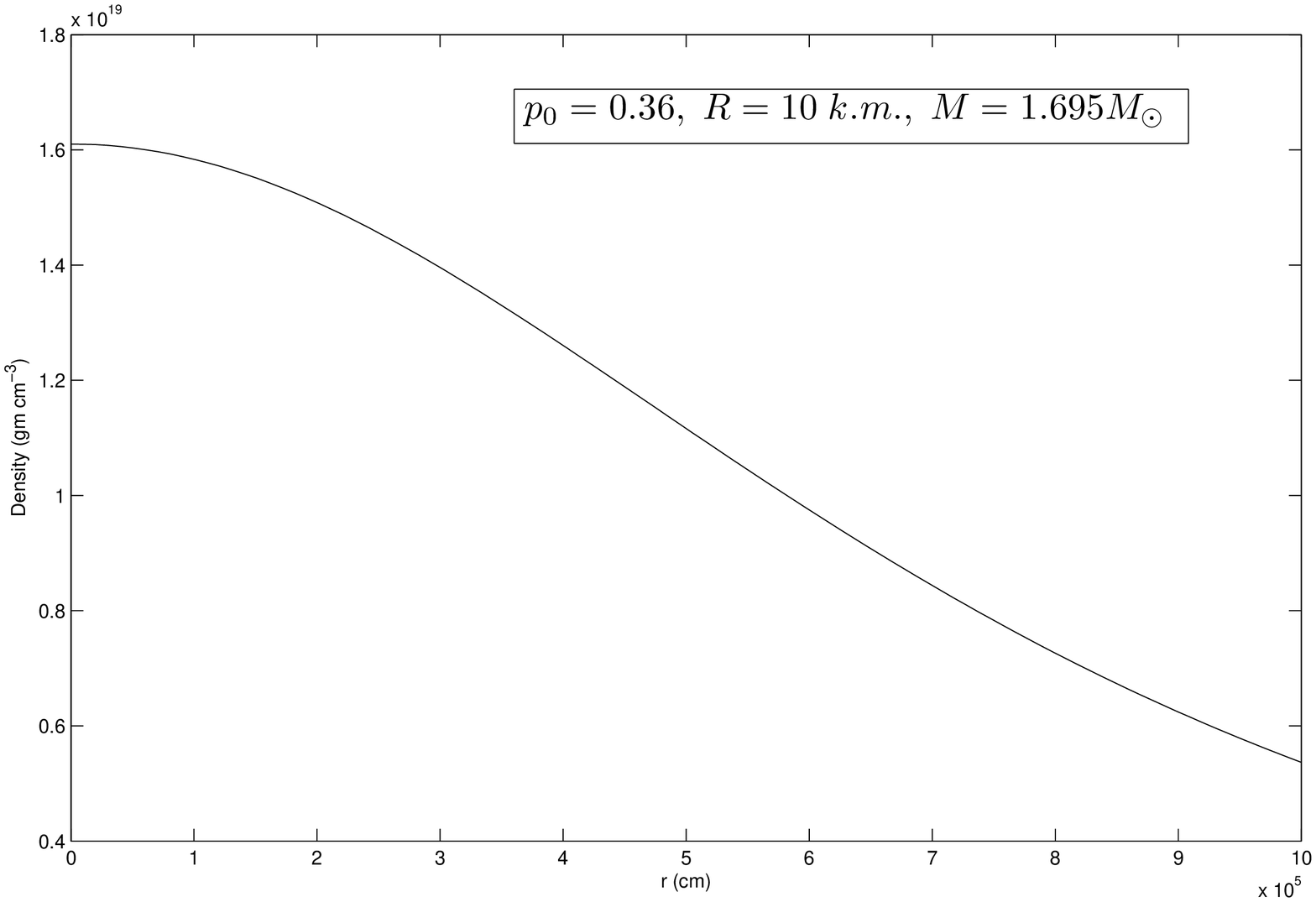,width=14cm}}
\vspace*{12pt}
\caption{Variation of density ($\rho$) against the radial parameter. ($1~$MeV~fm$^{-3} = 1.78\times 10^{12}~$gm~cm$^{-3}$).}
\end{figure}

\begin{figure}[pb]
\label{fig:2}
\centerline{\psfig{file=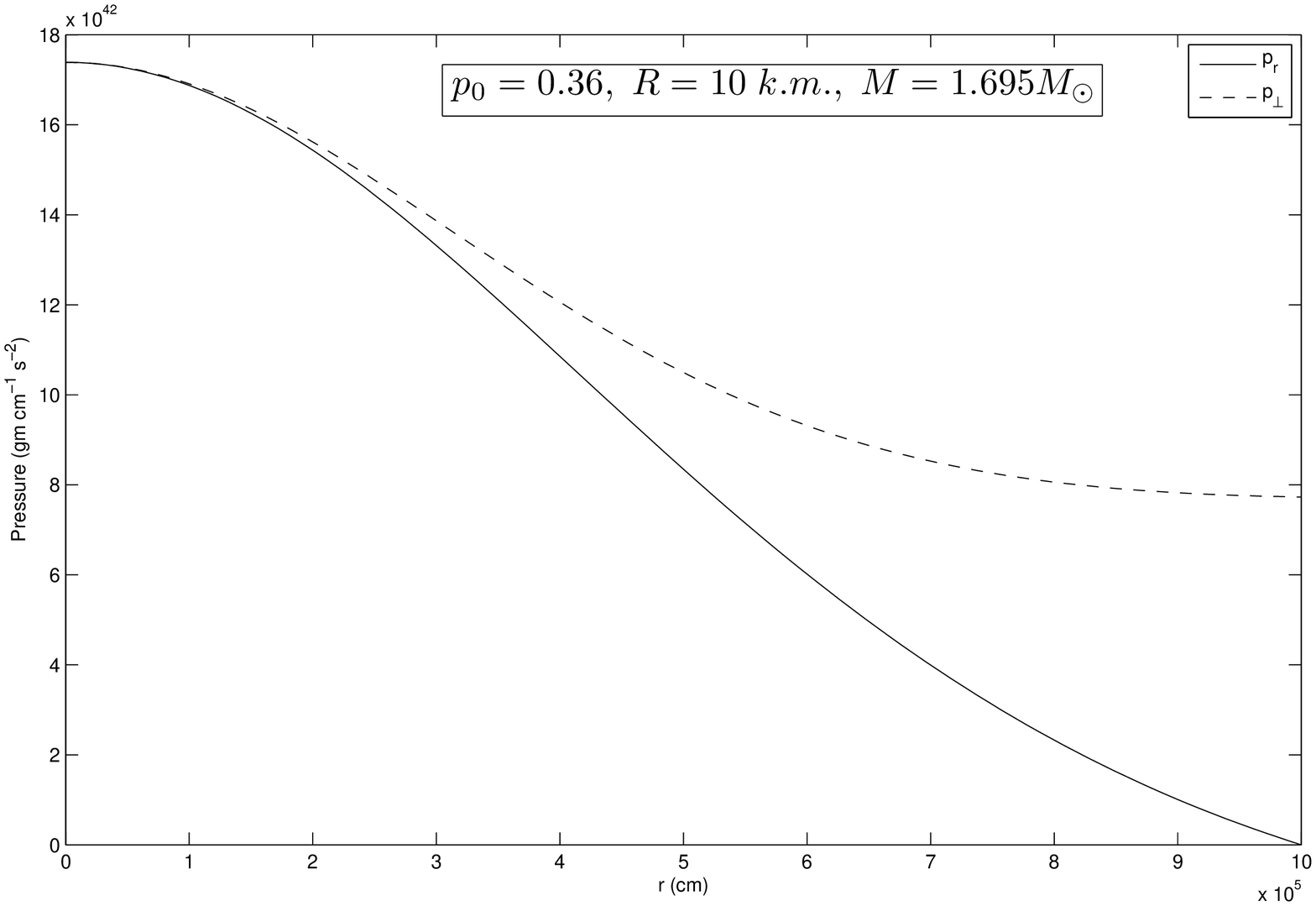,width=14cm}}
\vspace*{12pt}
\caption{Variation of pressure ($p_r$ and $p_{\perp}$) against the radial parameter  $r$.}
\end{figure}

\begin{figure}[pb]
\label{fig:3}
\centerline{\psfig{file=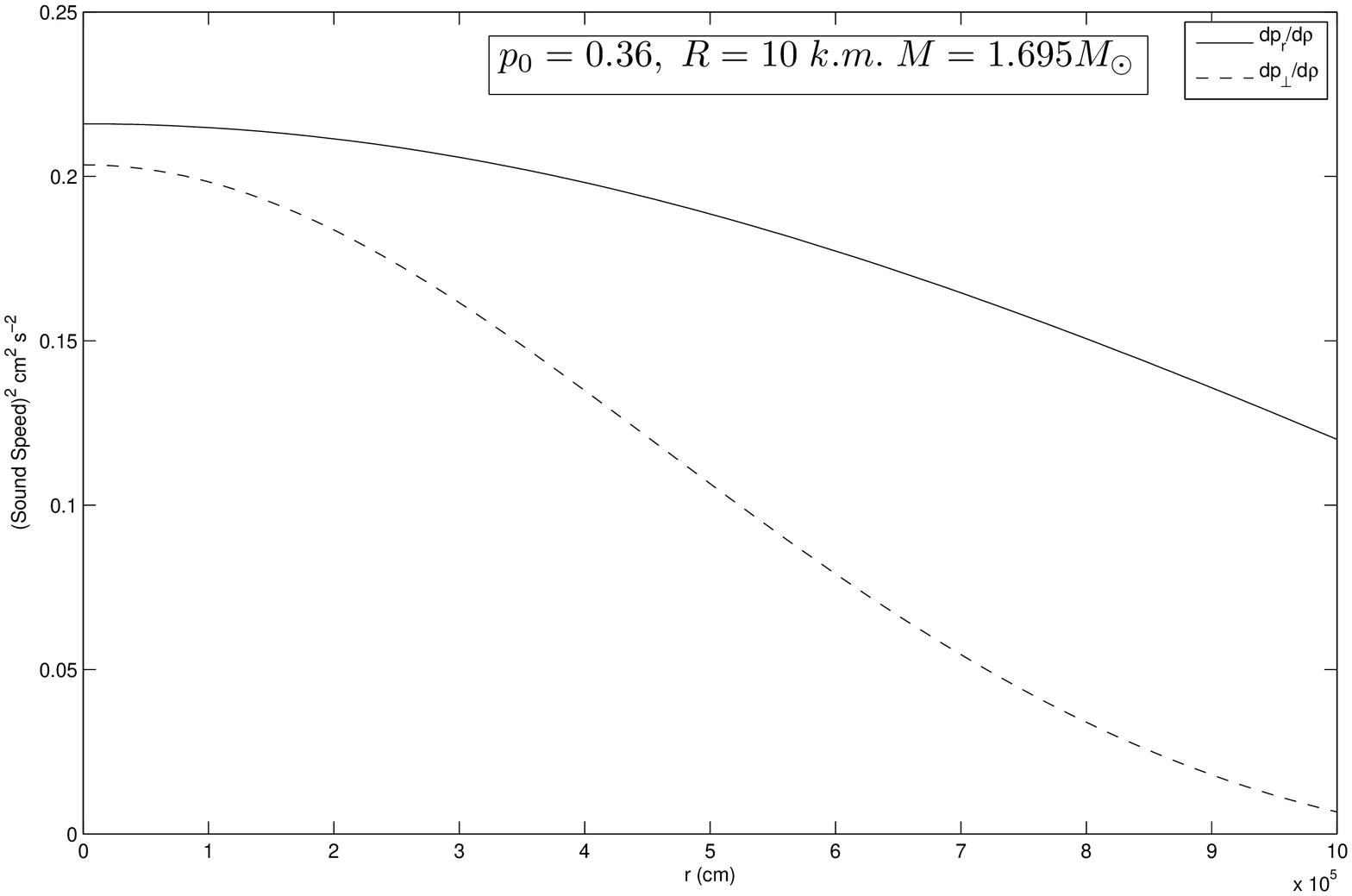,width=12cm}}
\vspace*{12pt}
\caption{Variation of $\frac{dp_r}{d\rho}$ against the radial parameter $r$.}
\end{figure}

\begin{figure}[pb]
\label{fig:4}
\centerline{\psfig{file=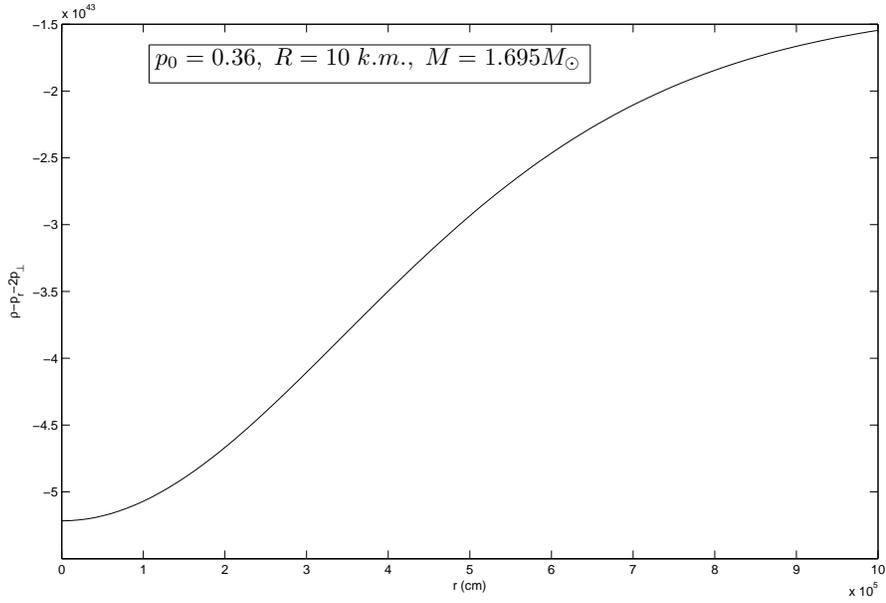,width=14cm}}
\vspace*{12pt}
\caption{Variation of $\rho -p_r -2p_{\perp}$ against the radial parameter $r$.\label{fig:5}}
\end{figure}

\begin{figure}[pb]
\label{fig:6}
\centerline{\psfig{file=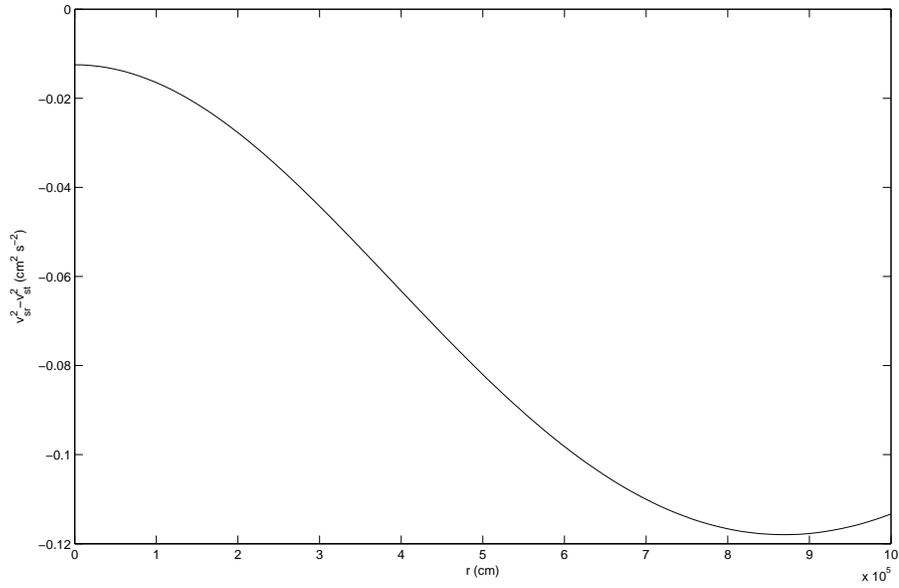,width=14cm}}
\vspace*{12pt}
\caption{Variation of $v_{sr}^2-v_{sp}^2$ against the radial parameter $r$.}
\end{figure}

We now analyze the gross behaviour of the physical parameters of our model such as energy density and two pressures at the interior of the star. For a particular choice $p_0 = 0.36$ (consistent with the bound), plugging in $c$ and $G$ at appropriate places, we have calculated the mass $M$, central density $\rho_c$ and surface density $\rho_R$ of a star of radius $R$. This has been shown in Table~$1$. We note that the central density in each case (except $VIII$, where we have assumed a comparatively larger radius which in turn has generated a bigger mass) lies above the deconfinement density\cite{Heinz,Karsch} $\sim 700~$MeV~fm$^{-3}$ which implies that quark phases may exist at the interiors of such configurations. Variations of the physical parameters for a particular case $VI$ have been shown in Fig.~(\ref{fig:1})-(\ref{fig:5}). The figures clearly indicate that the physical parameters are well-behaved and all the regularity conditions discussed above are satisfied at all interior points of the star. Moreover, the assumed parameters generate a stable configuration as shown in Fig.~(\ref{fig:6}). 

\begin{table}
\tbl{Values of the physical parameters for different radii with $p_0 =0.36$.}
{\begin{tabular}{@{}lcccc@{}} \toprule
Case & $R$ & $M$ & $\rho_{c}$  & $\rho_{R}$ \\
 & (km) & ($M_{\odot}$) & (MeV~fm$^{-3}$) & (MeV~fm$^{-3}$) \\ \colrule
I & 6.55 & 1.11 & 2108.46 & 702.82 \\
II & 6.7  & 1.14 & 2015.11 & 671.70 \\
III & 7.07 & 1.20 & 1809.71 & 603.24  \\
IV & 8    & 1.36 & 1413.41 & 471.14 \\
V & 9    & 1.53 & 1116.77 & 372.26 \\
VI & 10   & 1.69 & 904.58  & 301.53  \\
VII & 11   & 1.86 & 747.59  & 249.20 \\
VIII & 12   & 2.03 & 628.18  & 209.39 \\ \botrule
\end{tabular} \label{tab1}}
\end{table}

\subsection{Generating approximated EOS}
\begin{figure}[pb]
\label{fig:7}
\centerline{\psfig{file=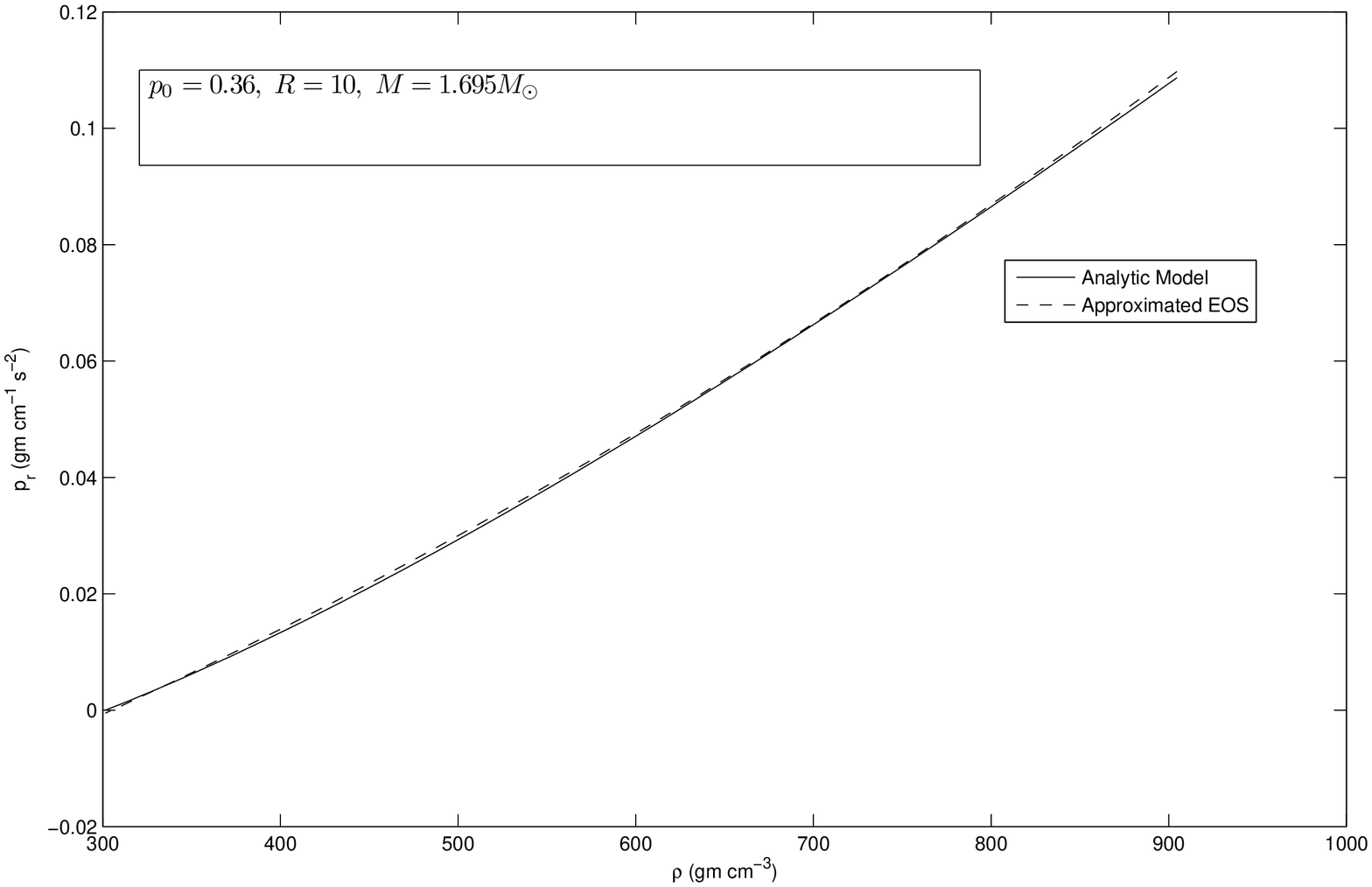,width=14cm}}
\vspace*{12pt}
\caption{Equation of state (EOS) generated from the analytic model (solid line) has been shown to be in agreement with the assumed quadratic EOS (dashed line).}
\end{figure}

Having derived a physically acceptable model, question to be asked is, what kind of material composition can be predicted for the stellar configurations admissible in this model? In other words, what would be the EOS corresponding to the material compositions of the configurations constructed from the model? Though construction of an EOS is essentially governed by the physical laws of the system, one can parametrically relate the energy-density and the radial pressure from the mathematical model which may be useful in predicting the composition of the system.  Making use of Eqs.~(\ref{eq16}) and (\ref{eq19}), we have plotted variation of the radial pressure against the energy-density as shown by the solid curve in Fig.~(\ref{fig:7}). Our intention now is to prescribe an approximate EOS which can produce similar kind of curve. Though, in principle, a barotropic EOS ($p_r=p_r(\rho)$) can be generated from Eqs.~(\ref{eq16}) and (\ref{eq19}) by eliminating $r$, we assume that the relevant EOS has the form
\begin{equation}
p_{r} = \rho_0 +\alpha\rho + \beta\rho^2,\label{eq41}
\end{equation}
where $\rho_0$, $\alpha$ and $\beta$ are constants. We make use of this EOS to plot $\rho$ vs $p_r$ which turns out to be almost identical to the curve generated from the analytic model if we set $\rho_0 = - 0.36$, $\alpha = 9.6 \times 10^{-5}$ and $\beta = 7.2 \times 10^{-8}$ (dashes curve in Fig.~(\ref{fig:7})). Though this has been shown to be true for a particular choice (case $VI$), it can be shown that the model admits the quadratic EOS (\ref{eq41}) for different choices of the parameters as well.

\section{Discussion}
Making use of Finch and Skea\cite{Finch} ansatz, we have generated exact solutions of Einstein's field equations representing a static spherically symmetric anisotropic stellar configuration. Bounds on the model parameters have been obtained on physical grounds and it has been shown that model is stable for $\frac{1}{3} < p_{0} < 0.3944$. Note that $p_0/R^2$ denotes the central density in this model and, therefore, the bound indicates that for a given radius or mass arbitrary choice of the central density is not permissible in this model. We have shown that the model admits an EOS which is quadratic in nature. Mathematically, this may be understood in the following manner. The ansatz (\ref{eq15}), together with the assumption (\ref{eq19}), generates an anisotropic stellar model whose composition may be described by the EOS of the form (\ref{eq41}). Note that in Ref.~\refcite{Feroze,MT}, quadratic EOS have been assumed a priori to obtain exact solutions of Einstein's field equations. In this paper, we have shown that such an assumption is consistent with an analytical model which has been constructed by making use of the Finch and Skea\cite{Finch} ansatz having a clear geometrical representation. In cosmology, for an accelerating universe, a non-linear quadratic EOS has been shown to be relevant for the description of dark energy and dark matter\cite{Ananda}. What type of matter can generate such an EOS in the high density regime of an astrophysical object is a matter of further investigation and will be taken up elsewhere.  

\section*{Acknowledgments}
RS gratefully acknowledges support from the Inter-university Centre for Astronomy and Astrophysics (IUCAA), Pune, India, where a part of this work was carried out under its Visiting Research Associateship Programme. BSR is grateful to IUCAA, Pune, India, for providing facilities where the part of work was done. BSR also thanks V. O. Thomas for useful discussions.


\pagebreak

\begin{figure}[pb]
\centerline{\psfig{file=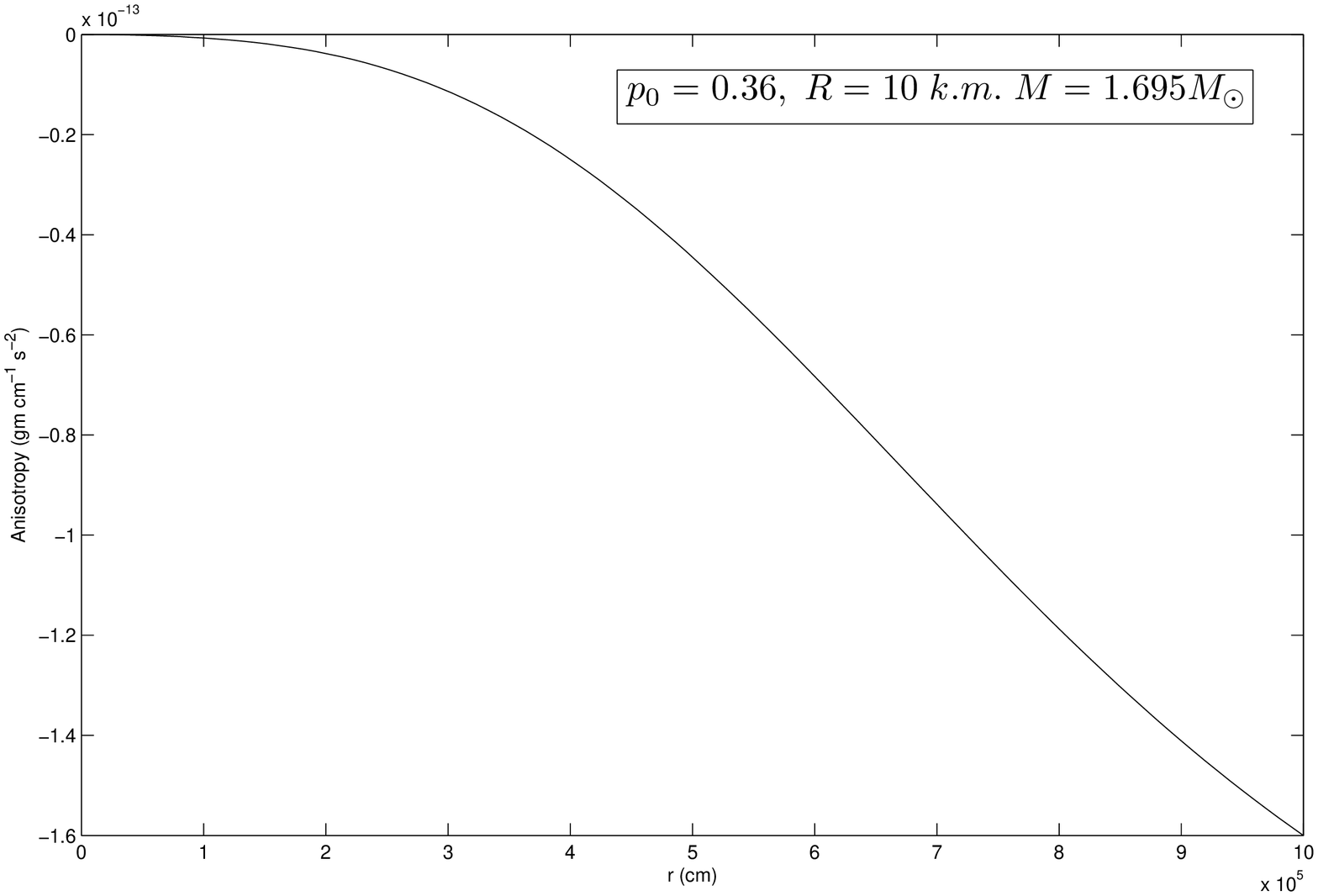,width=14cm}}
\vspace*{12pt}
\caption{Variation of anisotropic parameter$S(r)$ against the radial parameter $r$.}
\end{figure}

\end{document}